\documentclass[12pt,aps,superscriptaddress,onecolumn,nofootinbib,floatfix]{revtex4}

\usepackage{mathrsfs}
\usepackage{amsfonts,amssymb,amsmath}
\usepackage{graphicx,epsfig}
\usepackage{multirow}
\usepackage{verbatim}

\def\fsu5{$\cal{F}$-$SU(5)$}
\def\bfsu5{$\boldsymbol{\mathcal{F}}$-$\boldsymbol{SU(5)}$}

\def\m1half{$M_{1/2}$}
\def\m3half{$M_{3/2}$}
\def\m32{$M_{32}$}

\def\mt2{$M_{T2}$}
\def\x2{$\chi^2$}
\def\2b{$M_{T2}b$}

\def\bs0{$B_S^0 \rightarrow \mu^+ \mu^-$}

\begin{document}

\title{No-Scale Supergravity in the Light of LHC and Planck}

\author{Tianjun Li}

\affiliation{State Key Laboratory of Theoretical Physics and Kavli Institute for Theoretical Physics China (KITPC), Institute of Theoretical Physics, Chinese Academy of Sciences, Beijing 100190, P. R. China}

\affiliation{George P. and Cynthia W. Mitchell Institute for Fundamental Physics and Astronomy, Texas A$\&$M University, College Station, TX 77843, USA}

\author{James A. Maxin}

\affiliation{Department of Physics and Astronomy, Ball State University, Muncie, IN 47306 USA}

\author{Dimitri V. Nanopoulos}

\thanks{\normalsize Contribution to the Proceedings of the International School of Subnuclear Physics - 51st, Reflections on the Next Step for LHC, Erice, Sicily, Italy, 24 June - 3 July 2013, based on a talk given by Dimitri V. Nanopoulos.}

\affiliation{George P. and Cynthia W. Mitchell Institute for Fundamental Physics and Astronomy, Texas A$\&$M University, College Station, TX 77843, USA}

\affiliation{Astroparticle Physics Group, Houston Advanced Research Center (HARC), Mitchell Campus, Woodlands, TX 77381, USA}

\affiliation{Academy of Athens, Division of Natural Sciences, 28 Panepistimiou Avenue, Athens 10679, Greece}

\author{Joel W. Walker}

\affiliation{Department of Physics, Sam Houston State University, Huntsville, TX 77341, USA}


\begin{abstract}

We review the No-Scale \fsu5 model with extra TeV-scale vector-like flippon multiplets and associated collider phenomenology relating to the search for supersymmetry at the LHC. The model framework possesses the rather unique capacity to provide a light CP-even Higgs boson mass in the favored 124--126 GeV window while simultaneously retaining a testably light Supersymmetry (SUSY) spectrum. We further elucidate the emerging empirical connection between recent Planck satellite data and No-Scale Supergravity cosmological models that mimic the Starobinsky model of inflation,
drawing upon work conducted by the conference presenter (DVN) with John Ellis (King's College London/CERN) and Keith Olive (Minnesota).

\end{abstract}



\preprint{ACT-10-13}

\maketitle


\section{Introduction}

We have developed a framework named No-Scale \fsu5~\cite{
Li:2010ws, Li:2010mi,Li:2010uu,Li:2011dw, Li:2011hr, Maxin:2011hy,
Li:2011xu, Li:2011in,Li:2011gh,Li:2011rp,Li:2011fu,Li:2011ex,Li:2011av,
Li:2011ab,Li:2012hm,Li:2012tr,Li:2012ix,Li:2012yd,Li:2012qv,Li:2012jf,Li:2012mr,Li:2013hpa,Li:2013naa,Li:2013bxh,Li:2013mwa}
that is based upon the tripodal foundations of
the dynamically established boundary conditions
of No-Scale Supergravity, the Flipped $SU(5)$ Grand
Unified Theory (GUT), and a pair of TeV-scale
hypothetical ``{\it flippon}'' vector-like
super-multiplets motivated within local F-theory model
building.  The union of these features has
been demonstrated to naturally resolve a number
of standing theoretical problems, and to compare
favorably with experimental observation of the real world.

The aggressively minimalistic formalism of No-Scale
Supergravity~\cite{Cremmer:1983bf,Ellis:1983sf, Ellis:1983ei, Ellis:1984bm, Lahanas:1986uc}
provides for a deep connection to string theory in the infrared limit,
the natural incorporation of general coordinate invariance (general relativity),
a mechanism for Supersymmetry (SUSY) breaking which preserves a vanishing cosmological constant at the tree level
(facilitating the observed longevity and cosmological flatness of our Universe~\cite{Cremmer:1983bf}),
natural suppression of CP violation and flavor-changing neutral currents, dynamic stabilization
of the compactified spacetime by minimization of the loop-corrected scalar potential, and a dramatically
parsimonious reduction in parameterization freedom.
The split-unification structure of flipped $SU(5)$~\cite{Nanopoulos:2002qk,Barr:1981qv,Derendinger:1983aj,Antoniadis:1987dx}
provides for fundamental GUT scale Higgs representations (not adjoints), natural doublet-triplet
splitting, suppression of dimension-five proton decay~\cite{Antoniadis:1987dx,Harnik:2004yp},
and a two-step see-saw mechanism for neutrino masses~\cite{Ellis:1992nq,Ellis:1993ks}.
Modifications to the one-loop gauge $\beta$-function coefficients $b_i$ induced by inclusion of the
vector-like flippon multiplets create an essential flattening of the $SU(3)$ Renormalization Group Equation (RGE)
running ($b_3 = 0$)~\cite{Li:2010ws}, which translates into a wide separation between the primary
$SU(3)_C \times SU(2)_L$ unification near $10^{16}$~GeV and the secondary $SU(5) \times U(1)_X$ unification
near the Planck mass.  The corresponding baseline extension for logarithmic running of the
No-Scale boundary conditions, especially that imposed ($B_\mu = 0$) on the soft SUSY breaking term $B_\mu$
associated with the Higgs bilinear mass mixing $\mu$, allows sufficient room for natural dynamic
evolution into phenomenologically viable values at the electroweak scale.  Associated flattening
of the color-charged gaugino mass scale generates a stable characteristic mass texture of
$M(\widetilde{t}_1) < M(\widetilde{g}) < M(\widetilde{q})$, featuring a light stop and
gluino that are lighter than all other squarks~\cite{Li:2011ab}.

At this vital juncture in the LHC's history, entering an operational
pause for refitting to enable collisions at the full design energy of 14~TeV,
we take stock of the No-Scale \fsu5 model's full experimental status and prospects.
The first phase of LHC data collection at $\sqrt{s} = 7,8$~TeV
was highly noteworthy for both a discovery (the isolation
of a Standard Model (SM) like Higgs boson at around 125~GeV~\cite{:2012gk,:2012gu,Aaltonen:2012qt})
and a null result (the stubborn persistence of SUSY to unambiguously rise above the noise floor with increasing
luminosity and energy).  Both circumstances have pressed
the standard Constrained Minimal Supersymmetric Standard Model (CMSSM)
and Minimal Supergravity (mSUGRA) parameter spaces to an extreme~\cite{Strumia:2011dv,Baer:2012uya,Buchmueller:2012hv}.
As is natural, the \fsu5 model space has likewise diminished as
it has been probed by incoming collider data~\cite{Li:2013hpa}, although
a wealth of sufficient room to maneuver remains. This result is made all
the more remarkable by the fact that the No-Scale \fsu5 construction
takes the form of a Minimal Parameter Model (MPM), with all essential
experimental characteristics established solely by the universal gaugino
mass boundary $M_{1/2}$, according to which the full SUSY particle
spectrum is proportionally rescaled {\it en masse}. A MPM of similar form has been much studied in the past (See~\cite{Lopez:1995hg} and references therein. For a review, see~\cite{superworld}). In particular, the precise nature of the high-mass cutoff enforced on the model is at the point where the
charged stau and neutral lightest supersymmetric particle (LSP)~\cite{Ellis:1983ew} mass degeneracy
becomes so tight that the Planck~\cite{Ade:2013zuv} and WMAP~\cite{Spergel:2003cb,Spergel:2006hy,Komatsu:2010fb,Hinshaw:2012aka} cold dark matter (CDM)
relic density observations cannot be satisfied.  This hard upper boundary
on the model's leading dimensionful parameter, $M_{1/2}$, constitutes a top-down theoretical
mandate for a comparatively light (and testable) SUSY spectrum which does not excessively
stress natural resolution of the gauge hierarchy problem, and is in itself
a rather distinctive and unique No-Scale \fsu5 characteristic.

A recent analysis~\cite{Ellis:2013xoa,Ellis:2013nxa,Ellis:2013nka}
suggests that a cosmological model based upon the No-Scale supergravity sector yields compatibility
with the Planck satellite measurements. With convenient superpotential parameter choices, the new
cosmological model compatible with Planck data is a No-Scale supergravity realization of the
Starobinsky model of inflation~\cite{Starobinsky:1980te,Mukhanov:1981xt,Starobinsky:1983zz}. We shall elaborate here upon this intriguing connection between the No-Scale \fsu5 GUT model and a No-Scale Wess-Zumino model of inflation.

\section{The No-Scale \fsu5 Model}

Supersymmetry naturally solves
the gauge hierarchy problem in the SM, and suggests (given $R$ parity conservation)
the LSP as a suitable cold dark matter candidate.
However, since we do not see mass degeneracy of the superpartners,
SUSY must be broken around the TeV scale. In GUTs with
gravity mediated supersymmetry breaking, called
the supergravity models,
we can fully characterize the supersymmetry breaking
soft terms by four universal parameters
(gaugino mass $M_{1/2}$, scalar mass $M_0$, trilinear soft term $A$, and
the low energy ratio of Higgs vacuum expectation values (VEVs) $\tan\beta$),
plus the sign of the Higgs bilinear mass term $\mu$.

No-Scale Supergravity was proposed~\cite{Cremmer:1983bf}
to address the cosmological flatness problem,
as the subset of supergravity models
which satisfy the following three constraints:
i) the vacuum energy vanishes automatically due to the suitable
 K\"ahler potential; ii) at the minimum of the scalar
potential there exist flat directions that leave the
gravitino mass $M_{3/2}$ undetermined; iii) the quantity
${\rm Str} {\cal M}^2$ is zero at the minimum. If the third condition
were not true, large one-loop corrections would force $M_{3/2}$ to be
either identically zero or of the Planck scale. A simple K\"ahler potential that 
satisfies the first two conditions is~\cite{Ellis:1984bm,Cremmer:1983bf}
\begin{eqnarray} 
K &=& -3 {\rm ln}( T+\overline{T}-\sum_i \overline{\Phi}_i
\Phi_i)~,~
\label{NS-Kahler}
\end{eqnarray}
where $T$ is a modulus field and $\Phi_i$ are matter fields, which parameterize the non-compact $SU(N,1)/SU(N) \times U(1)$ coset space.
The third condition is model dependent and can always be satisfied in
principle~\cite{Ferrara:1994kg}.
For the simple K\"ahler potential in Eq.~(\ref{NS-Kahler})
we automatically obtain the No-Scale boundary condition
$M_0=A=B_{\mu}=0$ while $M_{1/2}$ is allowed,
and indeed required for SUSY breaking.
Because the minimum of the electroweak (EW) Higgs potential
$(V_{EW})_{min}$ depends on $M_{3/2}$,  the gravitino mass is 
determined by the equation $d(V_{EW})_{min}/dM_{3/2}=0$.
Thus, the supersymmetry breaking scale is determined
dynamically. No-Scale supergravity can be
realized in the compactification of the weakly coupled
heterotic string theory~\cite{Witten:1985xb} and the compactification of
M-theory on $S^1/Z_2$ at the leading order~\cite{Li:1997sk}.

In order to achieve true string-scale gauge coupling unification
while avoiding the Landau pole problem,
we supplement the standard ${\cal F}$-lipped $SU(5)\times U(1)_X$~\cite{Nanopoulos:2002qk,Barr:1981qv,Derendinger:1983aj,Antoniadis:1987dx}
SUSY field content with the following TeV-scale vector-like multiplets (flippons)~\cite{Jiang:2006hf}
\begin{eqnarray}
\hspace{-.3in}
& \left( {XF}_{\mathbf{(10,1)}} \equiv (XQ,XD^c,XN^c),~{\overline{XF}}_{\mathbf{({\overline{10}},-1)}} \right)\, ,&
\nonumber \\
\hspace{-.3in}
& \left( {Xl}_{\mathbf{(1, -5)}},~{\overline{Xl}}_{\mathbf{(1, 5)}}\equiv XE^c \right)\, ,&
\label{z1z2}
\end{eqnarray}
where $XQ$, $XD^c$, $XE^c$, $XN^c$ have the same quantum numbers as the
quark doublet, the right-handed down-type quark, charged lepton, and
neutrino, respectively.
Such kind of models can be realized in ${\cal F}$-ree ${\cal F}$-ermionic string
constructions~\cite{Lopez:1992kg},
and ${\cal F}$-theory model building~\cite{Jiang:2009zza,Jiang:2009za}. Thus, they have been 
dubbed ${\cal F}$-$SU(5)$~\cite{Jiang:2009zza}.

\section{No-Scale Supergravity and the Starobinsky Model~\cite{Ellis:2013xoa,Ellis:2013nxa,Ellis:2013nka}}

Recently, an added phenomenological boost has been given to No-Scale Supergravities by detailed
measurement of the Cosmic Microwave Background (CMB) perturbations (the structural seeds of galactic
supercluster formation residually imprinted upon the faint afterglow of the big bang) from the Planck~\cite{Ade:2013uln} satellite.  This experiment verified a highly statistically significant tilt $n_s < 1$ in the spectrum of scalar perturbations, and set stronger upper limits on the ratio $r$ of tensor (directional) to
scalar (isotropic) perturbations.  These measurements, particularly of $n_s$, place many leading
models of cosmic inflation in jeopardy.
For example, single-field models with a monomial potential $\phi^n: n \ge 2$
are now disfavored at the $\sim 95$\% CL in the case of $\phi^2$ models, and at higher CLs for
models with $n > 2$. This has revived interest in non-monomial single-field potentials, such as
that found in the minimal Wess-Zumino model~\cite{Croon:2013ana}~\footnote{Models with similar potentials
were proposed  long ago~\cite{Linde:1984cd,Linde1,Albrecht:1984qt} and more recently in~\cite{Kallosh:2007wm}: see~\cite{Olive:1989nu} for a review.}.

A curious scenario suggested by Starobinsky~\cite{Starobinsky:1980te}
in 1980 is known~\cite{Mukhanov:1981xt} to match the CMB data effortlessly,
yielding a value of $n_s \sim 0.96$ that is in perfect accord with experiment,
and a value of $r \sim 0.004$ that is comfortably consistent with the Planck upper limit~\cite{Ade:2013uln}.
This model is a rather ad-hoc modification of Einstein's description of gravity, which combines a quadratic power of the
Ricci scalar with the standard linear term.  At face value, this model is rather difficult to
take seriously, but there is substantial enthusiasm for the observation that this esoteric model is
in fact conformally equivalent to the low energy limit of No-Scale supergravity with a non-minimal $N_C \ge 2$
scalar sector~\cite{Ellis:2013xoa,Ellis:2013nxa}.  To be specific, the algebraic equations of motion corresponding
to a scalar field $\Phi$ with a quadratic potential that couples to a conventional Einstein term
may be freely substituted back into the action, resulting in the phenomenologically favorable
quadratic power of the scalar curvature~\cite{Stelle:1977ry,Whitt:1984pd}.

In considering the fundamental problem of how cosmological inflation fits into particle physics, a
point of view has been taken~\cite{Ellis:2013xoa,Ellis:2013nxa,Ellis:2013nka} that this union cries out for supersymmetry~\cite{Ellis:1982ed,Ellis:1982dg,Ellis:1982ws}, in
the sense that it requires an energy scale hierarchically smaller than the Planck
scale, thanks to either a mass parameter being $\ll M_P$ and/or a scalar self-coupling
being $\ll {\cal O}(1)$.  Since cosmology necessarily involves consideration of gravity,
it is natural to consider inflation in the context of local supersymmetry, i.e., supergravity~\cite{Freedman:1976xh,Deser:1976eh},
which points in turn to the superstring as a sole contender for the consistent master embedding of quantum gravity.
This preference is complicated, however, by the fact that a generic supergravity theory
has supersymmetry-breaking scalar masses of the same order as the gravitino mass, 
giving rise to the so-called $\eta$ problem~\cite{Copeland:1994vg,Stewart:1994ts} (Also see, for example, Refs.~\cite{Lyth:1998xn,Martin:2013tda}), where the large vacuum energy density
during inflation leads to masses for all scalars of order the Hubble parameter \cite{Goncharov:1984qm}.
While inflationary models in simple supergravity can be constructed 
to avoid the $\eta$ problem \cite{Nanopoulos:1982bv,Holman:1984yj}, these models rely on a seemingly
accidental cancellation, invoking extraneous fine tuning in the inflaton mass \cite{Linde:2007jn}.

For this reason, No-Scale supergravity has long been advocated~\cite{Cremmer:1983bf,Ellis:1983sf,Ellis:1983ei,Ellis:1984bm,Lahanas:1986uc} as the unique natural framework for
constructing models of inflation~\cite{Ellis:1984bf,Goncharov:1985ka,Binetruy:1987xj,Murayama:1993xu,Antusch:2009ty}, representing a low energy
limit of the superstring.  Moreover, this construction yields very successful low energy phenomenology, while invoking a bare
minimum (one or zero) of freely adjustable parameters.  These proposals have recently been reinvigorated in light of the
Planck data, constructing an SU(2,1)/SU(2) $\times$ U(1) No-Scale version of the minimal 
Wess-Zumino model~\cite{Ellis:2013xoa,Ellis:2013nxa,Ellis:2013nka}~\footnote{For an alternative supergravity incarnation of the
Wess-Zumino inflationary model, see~\cite{Nakayama:2013jka}.}.
It was shown that this NSWZ model is consistent with the Planck data for a
range of parameters that includes a special case in which it reproduces {\it exactly}
the effective potential and hence the successful predictions of the Starobinsky $R + R^2$ model~\cite{Ellis:2013xoa,Ellis:2013nxa,Ellis:2013nka}.

Starobinsky considered in 1980~\cite{Starobinsky:1980te} a generalization of the Einstein-Hilbert action to contain an $R^2$ contribution,
where $R$ is the scalar curvature:
\begin{equation}
S=\frac{1}{2} \int d^4x \sqrt{-g} (R+\alpha R^2) \, ,
\label{Staro}
\end{equation}
where $M \ll M_P$ is some mass scale. As was shown by Stelle in 1978~\cite{Stelle:1977ry} and by Whitt in 1984~\cite{Whitt:1984pd}, 
the theory (\ref{Staro}) is conformally equivalent to a theory combining canonical gravity with a scalar field $\varphi$,
described by
\begin{equation}
S=\frac{1}{2} \int d^4x \sqrt{-g} \left[(1 + 2\alpha \varphi) R - \alpha \varphi^2 \right] \, ,
\label{Whitt}
\end{equation}
as can be seen trivially using the Lagrange equation for $\varphi$ in (\ref{Whitt}).
Making the Weyl rescaling $\tilde{g}_{\mu\nu} = (1 + 2 \alpha \varphi) g_{\mu\nu}$,
equation (\ref{Whitt}) takes the form
\begin{equation}
S=\frac{1}{2} \int d^4x \sqrt{-g} \left[ R + \frac{6 \alpha^2 \partial^\mu \varphi \partial_\mu \varphi}
{(1 + 2 \alpha \varphi)^2} - \frac{\alpha \varphi^2}{(1 + 2 \alpha \varphi)^2} \right] \, .
\label{Cecotti4}
\end{equation}
Making now the field redefinition $\varphi^\prime = \sqrt{\frac{3}{2}} \ln \left( 1+ \frac{\varphi}{3 M^2} \right)$ with $\alpha = 1/6M^2$,
one obtains a scalar-field action with a canonical kinetic term:
\begin{equation}
S=\frac{1}{2} \int d^4x \sqrt{-\tilde{g}} \left[\tilde{R} + (\partial_\mu \varphi^\prime)^2 - \frac{3}{2} M^2 (1- e^{-\sqrt{2/3}\varphi^\prime})^2 \right] \, ,
\end{equation}
in which the scalar potential takes the form
\begin{equation} 
V =  \frac{3}{4} M^2 (1- e^{-\sqrt{2/3}\varphi^\prime})^2 \, .
\label{r2pot}
\end{equation}
The spectrum of cosmological density perturbations found by using (\ref{Staro})
for inflation were calculated by Mukhanov and Chibisov in 1981~\cite{Mukhanov:1981xt} and by
Starobinsky in 1983~\cite{Starobinsky:1983zz}. The current data on cosmic microwave
background (CMB) fluctuations, in particular those from the Planck satellite~\cite{Ade:2013uln},
are in excellent agreement with the predictions of this $R + R^2$ model.

Some general features of the effective low-energy theory derived from a generic supergravity theory are recalled from Refs.~\cite{Ellis:2013xoa,Ellis:2013nxa,Ellis:2013nka}. Neglecting gauge interactions, which are inessential for our purposes, any such theory
is characterized by a K\"ahler potential $K(\phi_i,  \phi^*_j)$, which is a hermitian function
of the chiral fields $\phi_i$ and their conjugates $\phi^*_j$, and a superpotential $W(\phi_i)$,
which is a holomorphic function of the $\phi_i$, via the combination
$G \equiv K + \ln W + \ln W^*$.  The effective field theory contains a generalized kinetic energy term
\begin{equation}
{\cal L}_{KE} \; = \; K^{ij^*} \partial_\mu \phi_i \partial \phi^*_j \, ,
\label{LK}
\end{equation}
where the K\"ahler metric $K^{ij^*} \equiv \partial^2 K / \partial \phi_i \partial \phi^*_{j}$, and the
effective scalar potential is
\begin{equation}
V \; = \; e^G \left[ \frac{\partial G}{\partial  \phi_i} K_{ij^*}  \frac{\partial G}{\partial  \phi^*_j} - 3 \right] \, ,
\label{effpot}
\end{equation}
where $K_{ij^*}$ is the inverse of the K\"ahler metric. Inserting into Eq.~(\ref{effpot}) the Kahler potential of Eq.~(\ref{NS-Kahler}), for N=2 and using the Wess-Zumino Superpotential 

\begin{equation}
W=\frac{\widehat{\mu}}{2} \Phi^2 - \frac{\lambda}{3} \Phi^3,
\label{superpotential}
\end{equation}
with $\widehat{\mu} = \mu (c/3)^{1/2}$, $c = 2<{\rm Re}~T>$, and $\lambda = \mu/3$,
we get the potential for the real part of the inflaton (see Ref.~\cite{Ellis:2013xoa} for details):      

\begin{equation}
V = \mu^2 e^{-\sqrt{2/3x}} {\rm sinh}^2(x/\sqrt{6}).
\label{inflatonpotential}
\end{equation}
Clearly, the Starobinsky potential of Eq.~(\ref{r2pot}) is identical with the No-Scale WZ potential of Eq.~(\ref{superpotential})!

\section{The Wedge of Bare-Minimal Constraints}

\begin{figure*}[htp]
        \centering
        \includegraphics[width=0.85\textwidth]{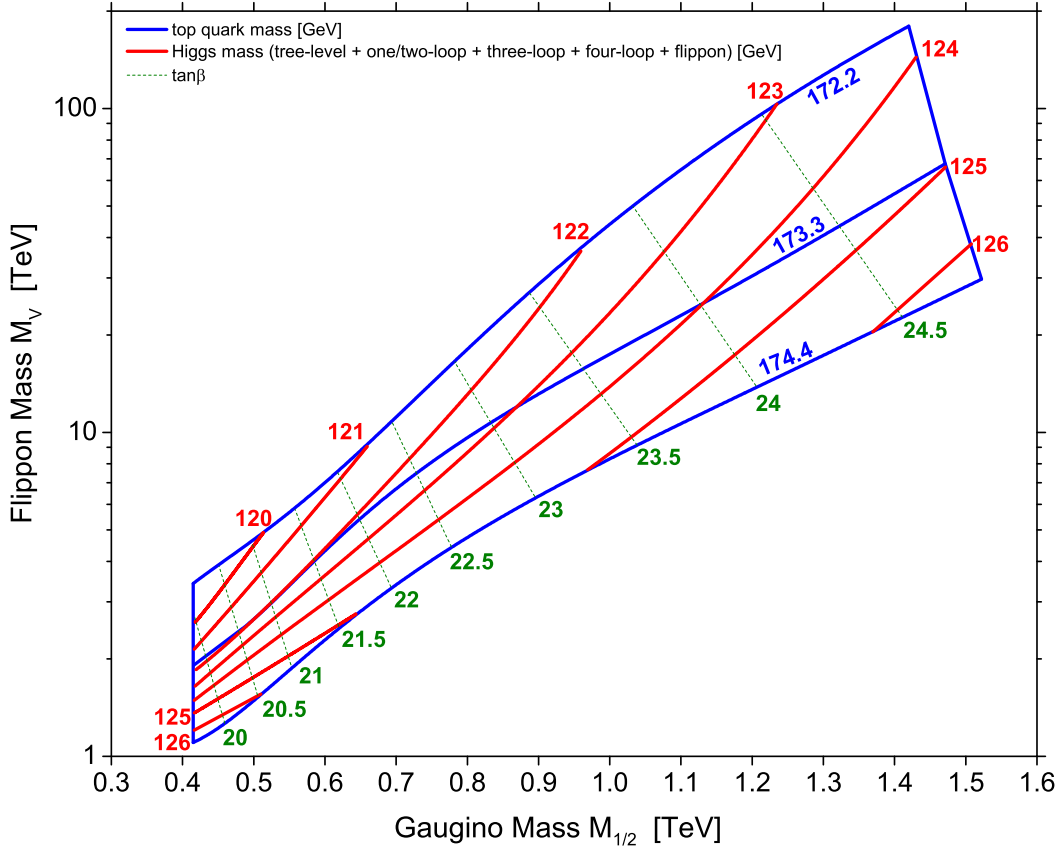}
        \caption{Constrained model space of No-Scale \fsu5 as a function of the gaugino mass $M_{1/2}$ and flippon mass $M_V$. The thick lines demarcate the total Higgs boson mass gradients, including the tree-level plus one/two-loop (as computed by the SuSpect~2.34 codebase), the three-loop plus four-loop contributions, and the flippon contribution. The thin dashed lines represent gradients of tan$\beta$, while the upper and lower exterior boundaries are defined by a top quark mass of $m_t = 173.3 \pm1.1$ GeV. The left edge is marked by the LEP constraints, while the right edge depicts where the Planck relic density can no longer be maintained due to an LSP and light stau mass difference less than the on-shell tau mass. All model space within these boundaries satisfy the Planck relic density constraint $\Omega h^2 = 0.1199 \pm 0.0027$ and the No-Scale requirement $B_{\mu}=0$.}
        \label{fig:higgstanb}
\end{figure*}

In Refs.~\cite{Li:2011xu,Li:2013naa}, we presented the wedge of No-Scale \fsu5
model space that is consistent with a set of ``bare minimal''
constraints from theory and phenomenology. The constraints included
i) consistency with the dynamically established boundary
conditions of No-Scale supergravity (most notably the
imposition of a vanishing $B_{\mu}$ at the final flipped $SU(5)$
GUT unification near $M_{\rm Pl}$, enforced as $\left|B_{\mu}\right(M_{\cal F})| \leq 1$ GeV, about
the size of the EW radiative corrections); ii) radiative electroweak
symmetry breaking; iii) the centrally observed WMAP7~\cite{Komatsu:2010fb} CDM relic density (and
now the Planck relic density $\Omega h^2 = 0.1199 \pm 0.0027$~\cite{Ade:2013zuv}) ; iv) the world average top-quark mass $m_t = 173.3 \pm 1.1$~GeV~\cite{:1900yx}; v) precision
LEP constraints on the light SUSY chargino and neutralino mass
content~\cite{LEP}; and vi) production of a lightest CP-even Higgs boson mass of $m_{h} = 125.5 \pm 1.5$
GeV, accomplished through additional tree level and one-loop contributions to the Higgs boson mass by
the flippon supermultiplets~\cite{Li:2011ab,Li:2012jf,Li:2013naa}, supplementing the Minimal
Supersymmetric Standard Model (MSSM) Higgs boson mass by just the essential additional 3-5 GeV amount
requisite to attain $m_{h} \sim 125$ GeV, while also preserving a testably light SUSY spectrum that does not reintroduce the gauge hierarchy problem via very heavy scalars that SUSY was originally intended to solve in the first place. This two-dimensional parameterization in the
vector-like {\it flippon} super-multiplet mass scale $M_V$
and the universal gaugino boundary mass scale $M_{1/2}$
was excised from a larger four-dimensional hyper-volume
also including the top quark mass $m_t$ and the ratio
$\tan \beta$. Surviving points, each capable
of maintaining the delicate balance required to satisfy 
$B_\mu = 0$ and the CDM relic density observations, were
identified from an intensive numerical scan, employing
MicrOMEGAs~2.1~\cite{Belanger:2008sj} to compute SUSY masses, using a proprietary modification of the
SuSpect~2.34~\cite{Djouadi:2002ze} codebase to run the
{\it flippon}-enhanced RGEs. The relic density, spin-independent cross-section, and all rare-decay process constraints have been computed with MicrOMEGAs~2.4~\cite{Belanger:2010gh}.

The union of all such points was found to consist of
a diagonal wedge ({\it cf.} Ref.~\cite{Li:2011xu,Li:2013naa}) in the $M_{1/2}$-$M_V$
plane, the width of which ({\it i.e.} at large $M_{1/2}$ and
small $M_V$ or vice-versa) is bounded by the central 
experimental range of the top quark mass, and the extent
of which ({\it i.e.} at large $M_{1/2} \sim 1500$~GeV and large $M_V$) is
bounded by CDM constraints and the transition to a charged
stau LSP. This upper region of the model space corresponds to an exponentially elevated {\it flippon}
mass $M_V$, which may now extend into the vicinity of 100~TeV.  This
delineation of the bare-minimally constrained \fsu5 parameter
space, including the correlated values of $m_t$, $\tan \beta$ and
the light CP-even Higgs mass for each model point, is depicted in
Figure~\ref{fig:higgstanb}. One obvious concern associated with this circumstance is the appearance
of a new intermediate scale of physics, and a potentially new associated
hierarchy problem.  However, we remark that the vector-like {\it flippon}
multiplets are free to develop their own Dirac mass, and are not in
definite {\it a priori} association with the electroweak scale symmetry
breaking; We shall therefore not divert attention here 
to the mechanism of this mass generation, although plausible
candidates do come to mind.

The advent of substantial LHC collision data in
the SUSY search rapidly eclipsed the tentative low-mass
boundary set by LEP observations.  A substantive correlation
in the \fsu5 mass scale favored by low-statistics excesses
in a wide range of SUSY search channels, particularly lepton-inclusive searches, at both CMS and ATLAS was remarked upon by
our group~\cite{Li:2012mr,Li:2013hpa} just below $M_{1/2} \sim 800$~GeV.
However, a minority of search channels, particularly lepton-exclusive
squark and gluino searches with jets and missing energy~\cite{ATLAS-CONF-2012-109},
were found to yield limits on $M_{1/2}$ that are inconsistent
with this fit, and that exert some limited tension against the upper
$M_{1/2}$ boundary of the model wedge.  This tension is also
reflected in one generic limit of a multijet plus a single lepton SUSY search from the CMS Collaboration
that places the gluino heavier than about 1.3~TeV~\cite{CMS-SUS-13-007}.

\section{Conclusions}

We have considered the experimental status and prospects of the No-Scale \fsu5 model and No-Scale Supergravity in the light of results from the LHC and Planck Satellite. Given that no conclusive
evidence for light Supersymmetry has emerged from the $\sqrt{s} = 7, 8$~TeV collider
searches as of yet, we described here the precise nature
of the high-mass cutoff enforced on this model at the point where the
stau and neutralino mass degeneracy becomes so tight that cold dark matter
relic density observations cannot be satisfied.  This hard upper boundary
on the model's mass scale constitutes a top-down theoretical
mandate for a comparatively light (and testable) SUSY spectrum which does not excessively
stress natural resolution of the gauge hierarchy problem. The No-Scale \fsu5 model is consistent with the dynamically established boundary conditions of No-Scale supergravity, radiative electroweak
symmetry breaking, the centrally observed Planck cold dark matter relic density, the world average
top-quark mass, precision LEP constraints on the light SUSY chargino and neutralino mass
content, and production of a $125.5 \pm 1.5$ GeV lightest CP-even Higgs boson mass.

Building upon ample extant phenomenological motivation for No-Scale \fsu5, we discussed the potentially significant empirical support recently provided to cosmological models of inflation based upon No-Scale Supergravity by intrinsic Starobinsky-like conformance with the Planck measurements, for a suitable choice of superpotential parameters. The potential deep connection between the No-Scale \fsu5 and No-Scale Wess-Zumino models is certainly testable in the near future by the upcoming 14 TeV LHC and future higher precision CMB measurements. Whereas the landscape of supersymmetric models is replete with predictions requiring years more of massive data observations and significantly higher LHC beam energies, we emphasize here to the contrary that all the No-Scale \fsu5 predictions dwell just on the cusp of an experimentally significant discovery. Detection of such a signal of stringy origin by the LHC and complementary experiments measuring the CMB perturbations could reveal not just the flipped nature of the high-energy theory, but also shed light on the geometry of the hidden compactified six-dimensional manifold. Thus, the stakes could not be higher for potential identification of supersymmetric events at the LHC within a No-Scale \fsu5 construction, and naturally realizable inflation in our \fsu5 No-Scale $SU(N,1)$ framework, or the revelations more profound.


\begin{acknowledgments}
Dimitri Nanopoulos would like to thank Professor A. Zichichi for his warm hospitality at Erice. This research was supported in part by the DOE grant DE-FG03-95-Er-40917 (DVN) and by the Natural Science Foundation of China under grant numbers 10821504, 11075194, 11135003, and 11275246 (TL). We also thank Sam Houston State University for providing high performance computing resources. The conference presenter (DVN) gratefully acknowledges contributions to section III of this proceedings report by John Ellis (CERN) and Keith Olive (Minnesota).
\end{acknowledgments}


\bibliography{bibliography}

\end{document}